\documentclass[aps,prl,twocolumn,superscriptaddress,nofootinbib,floatfix]{revtex4-2}
\usepackage{amsmath,amssymb,bm}
\usepackage{graphicx}
\usepackage{amsmath}
\usepackage{tikz}
\usepackage[percent]{overpic}
\newcommand{\lpmb}[1]{%
	\setbox0=\hbox{$#1$}%
	\kern-.01em\copy0
	\kern-\wd0
	\kern.01em\copy0
	\kern-\wd0
	\kern-.01em
	\raise.01em\box0}
\usepackage{xcolor}
\usepackage[colorlinks=true,linkcolor=black,citecolor=black,urlcolor=black]{hyperref}

\newcommand{\bl}[1]{{\color{black}#1}}

\newcommand{\re}[1]{{\color{black}#1}}

\begin{document}

%\title{Axion Emission from Ion-mediated  Nonlinear Plasma Current Filamentation Instabilities }
\title{Axion Emission Controlled by Ion Motion in Relativistic Beam--Plasma Filamentation}

\author{Zhi-Bo Zhang}
\affiliation{Department of Physics, Shanghai Normal University, Shanghai 200234, China}
\author{Pei-Lun He}
%\affiliation{State Key Laboratory of Dark Matter Physics, Key Laboratory for Laser Plasmas (Ministry of Education) and School of Physics and Astronomy, Collaborative Innovation Center for IFSA (CICIFSA), Shanghai Jiao Tong University, Shanghai 200240, China}
\affiliation{School of Physics and Astronomy, Shanghai Jiao Tong University, Shanghai 200240, China}
\author{Zheng Gong}
\email{zgong92@itp.ac.cn}
\affiliation{Institute of Theoretical Physics, Chinese Academy of Sciences, Beijing 100190, China}
\author{Yue-Yue Chen}
\email{yueyuechen@shnu.edu.cn}
\affiliation{Department of Physics, Shanghai Normal University, Shanghai 200234, China}

\date{\today}

\begin{abstract}
\bl{The interaction between relativistic electron beams and plasma can generate magnetic fields with strengths of $\sim 10^4$ T far exceeding conventional magnets, which offers a promising environment for producing weakly coupled particles. Using particle-in-cell simulations incorporating an axion-emission model, we show that a 10-GeV electron beam traversing dense plasma emits a forward-collimated axion beam during the ion-mediated nonlinear stage of current filamentation instability. Although the electron-driven linear instability seeds the filaments, ion motion governs their subsequent expansion and coalescence. 
This evolution reinjects cold-core electrons into large-amplitude orbits in phase space, where they repeatedly cross the strong-field flanks of neighboring magnetic vortices, thereby enhancing axion emission.  
At a background-plasma density of $\sim10^{21}\,\mathrm{cm}^{-3}$, suppressing ion motion reduces the axion yield by a factor of 14.5. 
Our results demonstrate that ion dynamics play a crucial role in characterizing the emission of weakly coupled particles induced by relativistic filamentation instabilities, suggesting a new beam-driven method for probing the axion-electron interaction at accelerators.}
%and explaining potential axion origin in analogical astrophysical scenarios.
\end{abstract}

\maketitle

Axions and axion-like particles (ALPs) provide well-motivated extensions to the Standard Model and are prominent dark-matter candidates.
Axions stem from the Peccei-Quinn solution to the strong-CP problem, while more general ALPs arise in theories with broken global symmetries~\cite{PecceiQuinn1977, Weinberg1978, Wilczek1978, DFSZ1981, IrastorzaRedondo2018}.
%\gr{Many established searches probe the two-photon coupling through macroscopic magnetic fields, optical systems, or crystalline fields}~\cite{Sikivie1983,CAST2017,Zavattini2008PVLAS,Ballou2015OSQAR,Halliday2025XFEL}.
Established searches for the two-photon coupling include helioscope experiments based on axion--photon conversion in laboratory magnetic fields~\cite{Sikivie1983,CAST2017}, optical-polarimetry searches~\cite{Zavattini2008PVLAS},
%light-shining-through-wall experiments~\cite{Ballou2015OSQAR},
light-shining-through-wall experiments and stimulated-conversion proposals~\cite{Ballou2015OSQAR,Beyer2022LSW}, laser-driven parametric conversion proposals~\cite{Beyer2023Parametric}, and crystalline-field searches~\cite{Halliday2025XFEL}.
The electron coupling \(g_{ae}\), present at tree level in Dine-Fischler-Srednicki-Zhitnitsky (DFSZ) models~\cite{Zhitnitsky1980,DiLuzio2020}, offers a complementary laboratory channel. 
Conventional electron beam-dump searches constrain this coupling through visible downstream decays, so they are sensitive only when the boosted decay length falls within a finite experimental window~\cite{Konaka1986,Riordan1987,Bjorken1988, Bross1991,LiuMiller2017,Eberhart2025}. 
Active-dump missing-energy searches avoid this restriction by inferring the energy carried by an escaping boson, however, they require event-by-event calorimetry and stringent control of pileup and rare backgrounds~\cite{Andreev2021NA64}. 
Reactor experiments provide another controlled source, although geometric dilution limits their nearly isotropic flux~\cite{NEON2025}.
More recently, coherent crystal-based production through the axion-electron coupling has also been proposed, exploiting phase matching to achieve macroscopic enhancement~\cite{Bai2026MacroscopicAxion}.

Axions may couple to photons, electrons, or nucleons, with each interaction motivating a distinct experimental strategy. 
Here we consider direct axion radiation from relativistic electrons in strong electromagnetic fields. 
Strong-field calculations have established the nonlinear Compton-like emission of pseudoscalar ALPs in prescribed laser backgrounds~\cite{Gies2009,King2018PLB,
DillonKing2018EPJC}, and laser-accelerated or electron-laser schemes have made this \(g_{ae}\) channel experimentally concrete~\cite{Vacalis2025LAE,Chen2026SpinAxion}. 
Their accumulated exposure is governed not by the peak field alone, but also by the number of electrons sampling the focal field, the interaction duration, and the repetition rate. 
However, simultaneously achieving high pulse energy and a high repetition rate remains technically challenging for high-power laser systems~\cite{Li2023Petawatt,Mirzaie2024NCS}. 
By contrast, high-charge relativistic electron beams can exploit plasma-generated electromagnetic fields to induce strong-field radiation processes without an external high-power laser~\cite{Matheron2023BeamPlasma}.
Here, each electron bunch drives current filamentation and emits axions within the resulting self-generated fields, allowing production to be repeated at the accelerator repetition rate.
Related beam--plasma experiments have recently been proposed for FACET-II~\cite{E305_proposal,E336_proposal}.

The current-filamentation (or beam--Weibel) instability provides the self-organized field structure. Counter-streaming beam and return currents separate into filaments, generating strong transverse magnetic fields~\cite{Weibel1959,Fried1959,Bret2008PRL,Bret2010,SanMiguelClaveria2026QSA}. The linear instability is predominantly electron-driven. After saturation, however, ion motion becomes decisive, wherein magnetic pressure expands the plasma cavities and drives filament coalescence, increasing both the magnetic-field amplitude and the transverse spatial scale~\cite{Honda2000,BretDieckmann2008,Peterson2021,Mathiak2025,Schoenwaelder2026}.

\bl{Here, we demonstrate that this ion-mediated nonlinear filamentation stage %contributes majority of the axion emission. 
\re{accounts for most of the axion emission.}
As neighboring filaments coalesce, electrons leave the cold cores to cross a moving separatrix in their phase space. 
These reinjected electrons repeatedly traverse the strong-field flanks of adjacent magnetic vortices, dominating the total radiation. 
The emitted axions remain forward-collimated, yielding high geometric acceptance for a downstream active dump.
Moreover, their sub-GeV-to-GeV spectrum lies well within the high-energy regime where axion-induced external pair production is efficient. Above approximately \(10~\mathrm{MeV}\), pair production becomes the dominant \(g_{ae}\)-mediated detection channel, making the source particularly favorable for downstream calorimetry~\cite{AriasAragon2025Pair}.
%{\color{green}
%As Weibel-type filamentation is ubiquitou in weakly magnetized relativistic flows.
%In astrophysical plasmas it mediates collisionless shocks and amplifies magnetic fields~\cite{MedvedevLoeb1999,Spitkovsky2008Shock,Gong2023Slingshot,Gong2025SpinPlasmoids}. The same ion-mediated evolution identified here may therefore influence astrophysical electron-coupled particle emission beyond laboratory beam-plasma systems.
More broadly, Weibel-type filamentation is widespread in weakly magnetized astrophysical plasmas, where it contributes to collisionless shock formation and magnetic-field amplification~\cite{MedvedevLoeb1999,Spitkovsky2008Shock,Gong2023Slingshot,Gong2025SpinPlasmoids}. The ion-mediated nonlinear dynamics identified here may therefore also influence axion emission in such environments.
}

We investigate this mechanism with an extended EPOCH code~\cite{Arber:2015hc,Gong2021Retrieving} that incorporates a new axion-emission module. The axion-electron coupling is described by
\begin{equation}
  \mathcal{L}_{ae}  = i g_{ae} a \bar{\psi}_e \gamma_5 \psi_e,
  \label{eq:gae}
\end{equation}
where \(a\) and \(\psi_e\) denote the ALP and electron fields, respectively, \(\gamma_5\) is the chirality matrix, and \(g_{ae}\) is dimensionless.
Within the local constant field approximation~\cite{Nikishov1964,Nikishov1967}, the differential probability for an electron of energy $\varepsilon$ to emit an axion of energy $\omega_a$ is given by~\cite{Chen2026SpinAxion,He2026AxionSemiclassical,DillonKing2018EPJC,SupplementalMaterial1}
\small{
\begin{equation}
\begin{split}
\frac{dW_a}{d\omega_a dt} = C_a \Bigg\{ &-\left[\frac{\varepsilon\varepsilon'}{2\omega_a^2}\delta_m^2\right] \operatorname{IntK}_{1/3}(z_q^a) \\
&+ \left(1 + \frac{\varepsilon\varepsilon'}{\omega_a^2}\delta_m^2\right) K_{2/3}(z_q^a) \Bigg\},
%\frac{dW_a}{d\omega_a dt} = C_a \Bigg\{ &-\left[\frac{\varepsilon\varepsilon'}{2\omega_a^2}\delta_m^2\right] \operatorname{IntK}_{1/3}(z_a^q) + \left(1 + \frac{\varepsilon\varepsilon'}{\omega_a^2}\delta_m^2\right) K_{2/3}(z_a^q) \Bigg\},
\label{eq:axion_rate}
\end{split}
\end{equation}}where $C_a = \sqrt{3}\,g_{ae}^2\,\omega_a^2\,m_e^2 / (24\pi^2\,\varepsilon^3\,\varepsilon')$ and $z_q^a = 2\omega_a\left(1 + \varepsilon\varepsilon'\delta_m^2/\omega_a^2\right)^{3/2} / (3\chi_e\varepsilon')$, with $\delta_m = m_a/m_e$, \re{where $m_a$ is the axion mass.} Here $\varepsilon' = \varepsilon - \omega_a$ is the electron energy after emission, $\operatorname{IntK}_{1/3}(z) \equiv \int_z^\infty dz' K_{1/3}(z')$, and $K_\nu$ is the modified Bessel function of the second kind. For each electron macroparticle, the local strong-field parameter
$\chi_e \equiv (e\hbar/m_e^3c^4)\sqrt{-(F_{\mu\nu}p^\nu)^2}$
is evaluated from its instantaneous momentum and interpolated
electromagnetic fields, where $F_{\mu\nu}$ is the electromagnetic
field tensor, $p^\nu$ is the electron four-momentum, $m_e$ and $-e$
are the electron mass and charge, respectively, and $\hbar$ is the
reduced Planck constant.
Axion-emission events are sampled using a Monte Carlo algorithm. 
In the weak-coupling regime considered here, axion emission does not affect the collective plasma evolution, and the source yield scales as $N_a \propto g_{ae}^2$. This framework combines the self-consistent beam--plasma evolution with QED radiation and perturbative ALP emission~\cite{SupplementalMaterial1}.

\begin{figure}[t]
	\centering
	\begin{overpic}[width=\columnwidth]{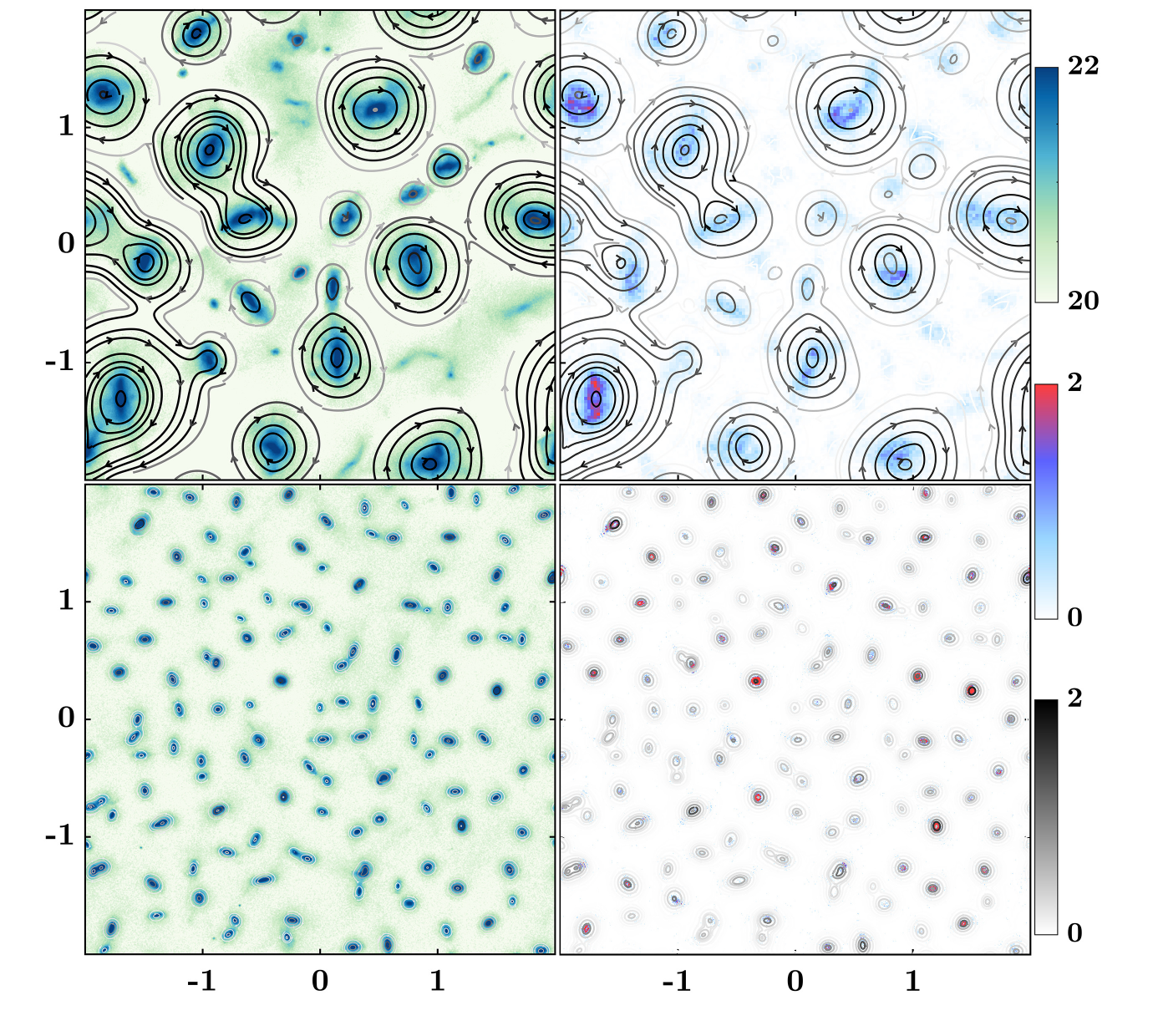}

\newcommand{\labelbox}[6]{%
  \put(#1,#2){%
    \makebox(0,0){%
      \tikz[baseline=(n.base)]{
        \node[
          fill=white,
          fill opacity=#6,
          text opacity=1,
          minimum width=#4,
          minimum height=#5,
          inner xsep=2pt,
          inner ysep=1pt,
          outer sep=0pt,
          rounded corners=2pt
        ] (n) {#3};
      }%
    }%
  }%
}
    
\labelbox{12}{86.5}{\textsf{(a)}}{2mm}{2mm}{0.85}
\labelbox{52}{86.5}{\textsf{(b)}}{2mm}{2mm}{0.85}
\labelbox{12}{45}{\textsf{(c)}}{2mm}{2mm}{0.85}
\labelbox{52}{45}{\textsf{(d)}}{2mm}{2mm}{0.85}
		
		%xylabel
		\put(26,0){\makebox(0,0){$\lpmb{x\,[\mu m]}$}}
		\put(68,0){\makebox(0,0){$\lpmb{x\,[\mu m]}$}}
		
		\put(1.5,68){\makebox(0,0){\rotatebox{90}{$\lpmb{y\,[\mu m]}$}}}
		\put(1.5,27){\makebox(0,0){\rotatebox{90}{$\lpmb{y\,[\mu m]}$}}}
		
		%colorbar
\put(98, 73.5){\makebox(0,0){\rotatebox{90}{$\lpmb{n_{be}\,[\lpmb{\mathrm{cm}^{-3}}]}$}}}
\put(98, 19.5){\makebox(0,0){\rotatebox{90}{$\lpmb{B_{\perp}}\,[\lpmb{10^{4}}\,\lpmb{\mathrm{T}}]$}}}
\put(98, 46.5){\makebox(0,0){\rotatebox{90}{$\lpmb{n_{a}}\,[\lpmb{\mathrm{A.U}}]$}}}

\labelbox{77}{86.5}{$\lpmb{\mathrm{Mobile\ ions}}$}{5mm}{4mm}{0.85}
\labelbox{76}{45}{$\lpmb{\mathrm{Immobile\ ions}}$}{5mm}{4mm}{0.85}

\labelbox{35}{86.5}{$\lpmb{\mathrm{Mobile\ ions}}$}{5mm}{4mm}{0.85}
\labelbox{34}{45}{$\lpmb{\mathrm{Immobile\ ions}}$}{5mm}{4mm}{0.85}
        
	\end{overpic}
	
\caption{Spatial distributions of (a) beam-electron density \(n_{be}\) and (b) axion-production density \(n_a\) at \(t=4~\mathrm{ps}\), where the streamlines denote the magnetic field \(B_\perp=(B_x^2+B_y^2)^{1/2}\). %Panels (c)(d) show the similar reresults but for the case with immobile ions. 
\re{Panels (c) and (d) show the corresponding
distributions for immobile ions.}
Mobile ions produce expanded and coalesced filament channels, whereas immobile ions retain narrow filaments.}
	\label{fig:source}
\end{figure}

\begin{figure}[t]
	\centering
	% Artwork update: replace B_{xy}, \varepsilon_B/\varepsilon_{be}, and
	% dN_{be}/d\varepsilon_{be} by B_\perp, U_B/U_{b0}, and dN_{be}/dE_e;
	% the n_{pe} color-bar unit should be 10^{29} m^{-3}, not cm^{-3}.
	
	%\includegraphics[width=\columnwidth]{axionfig1.eps}
	%\begin{overpic}[width=\columnwidth, grid, tics=10]{fig2_10.8.pdf}
	\begin{overpic}[width=\columnwidth]{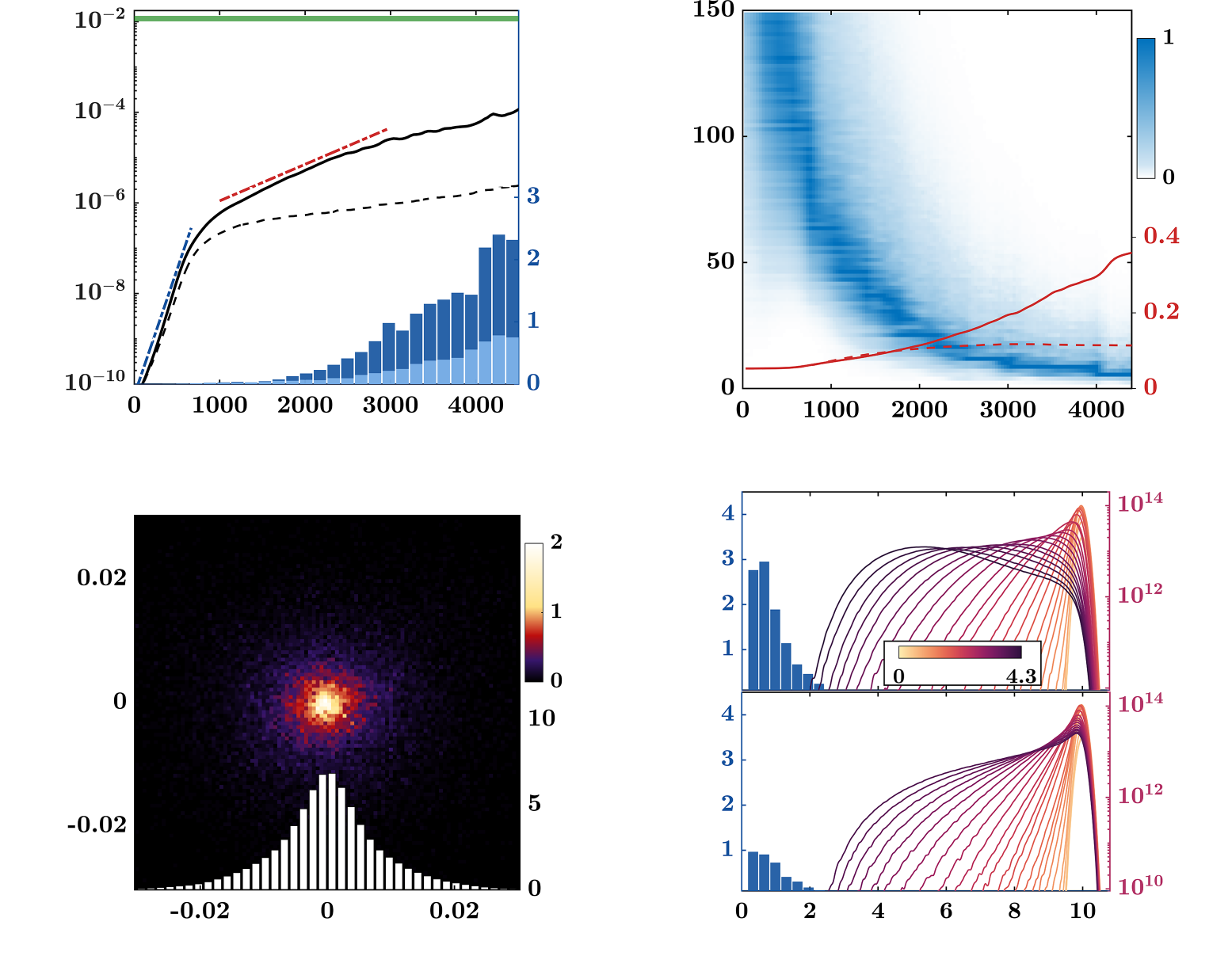}
		
		\definecolor{mylabelcolorblue}{rgb}{0.07, 0.32, 0.62} 
		\definecolor{mylabelcolorred}{rgb}{0.80, 0.16, 0.16} 
		\definecolor{mylabelcolorViolet}{rgb}{0.80, 0.16, 0.16}

		\put(12,32){\textcolor{white}{\textsf{(c)}}}
		\put(61,72){\textsf{(b)}}
		\put(12,72){\textsf{(a)}}
		\put(61,34){\textsf{(d)}}
		
		%fig a
	   \put(26,41){\makebox(0,0){\scriptsize$\lpmb{t}\,[\lpmb{\mathrm{fs}}]$}}
    
		\put(2.5,61){\makebox(0,0){\scriptsize\rotatebox{90}{$\lpmb{\varepsilon_B / \varepsilon_{be,0}}$}}}

        \put(47,55){\makebox(0,0){\scriptsize\rotatebox{90}{\textcolor{mylabelcolorblue}{$\lpmb{dN_a/dt\,[\mathrm{A.U}]}$}}}}

		\put(28,69){\makebox(0,0){\scriptsize\textcolor{mylabelcolorred}{$\lpmb{\varepsilon_B \propto e^{2\Gamma_{\rm nl}^{\rm fit}t}}$}}}
		\put(25,55){\makebox(0,0){\scriptsize\textcolor{mylabelcolorblue}{$\lpmb{\varepsilon_B \propto e^{2\Gamma_{\rm lin}^{\rm fit}t}}$}}}

		%fig b
	   \put(76,41){\makebox(0,0){\scriptsize$\lpmb{t}\,[\lpmb{\mathrm{fs}}]$}}
		\put(98,69){\makebox(0,0){\scriptsize\rotatebox{90}{$\lpmb{f(P)}$}}}
       \put(98,52){\makebox(0,0){\scriptsize\rotatebox{90}{\textcolor{mylabelcolorViolet}{$\lpmb{\lambda_P\,[\mu m]}$}}}}
		\put(54,61){\makebox(0,0){\scriptsize\rotatebox{90}{$\lpmb{K_\perp\,[\mu m^{-1}]}$}}}

	   \put(48,13){\makebox(0,0){\scriptsize\rotatebox{90}{$\lpmb{dN_{a}/d\theta_{x}}$}}}
		\put(48,28){\makebox(0,0){\scriptsize\rotatebox{90}{$\lpmb{g(n_a)}$}}}

        \put(26,1){\makebox(0,0){\scriptsize$\lpmb{\theta_x}\,[\lpmb{\mathrm{rad}}]$}}
        \put(2.5,20){\makebox(0,0){\scriptsize\rotatebox{90}{$\lpmb{\theta_y}\,[\lpmb{\mathrm{rad}}]$}}}

		\put(22.0,10){
			\tikz[baseline]{
				\draw[<->, line width=0.8pt, color=mylabelcolorred]
				(0,0) -- (0.545,0);
			}
		}
		\put(36.0,12.5){\makebox(0,0){\scriptsize\textcolor{white}{$\lpmb{\text{10 mrad}}$}}}
        
		%fig d
		\put(75.5,1){\makebox(0,0){\scriptsize$\lpmb{\varepsilon_{be},\varepsilon_a\,[\mathrm{GeV}]}$}}
        \put(77.8,23){\makebox(0,0){\tiny $\lpmb{t}\,[\lpmb{\mathrm{ps}}]$}}
		
        \put(97,21){\makebox(0,0){\scriptsize\rotatebox{90}{\textcolor{mylabelcolorViolet}{$\lpmb{dN_{be}/d\varepsilon_{be}\,[\lpmb{\mathrm{A.U}}]}$}}}}
        \put(55,20){\makebox(0,0){\scriptsize\rotatebox{90}{\textcolor{mylabelcolorblue}{$\lpmb{dN_a/d\varepsilon_a\,[\lpmb{\mathrm{A.U}}]}$}}}}

        \put(77,36){\makebox(0,0){\scriptsize $\lpmb{\mathrm{Mobile~ions}}$}}
        \put(77,20){\makebox(0,0){\scriptsize $\lpmb{\mathrm{Immobile~ions}}$}}
	\end{overpic}
	
\caption{\bl{(a) Time evolution of magnetic energy for mobile (solid) and immobile (dashed) ions. The green line indicates the nonlinear-saturation estimate $\varepsilon_B/\varepsilon_{be,0} \simeq (1/8)\min\{1,m_i/(\gamma_{b0}Z_i m_e)\}$~\cite{Peterson2021}. Blue and red dashed lines fit the linear and secondary ion-mediated growth, respectively.
(b) Normalized transverse beam-density-fluctuation power spectrum \(f(P)=P(K_\perp,t)/P_{\max}(t)\), where \(K_\perp=(k_x^2+k_y^2)^{1/2}\).
Red solid (dashed) curves show the power-weighted mean transverse scale \(\lambda_P=\langle 2\pi/K_\perp\rangle_P\) for mobile (immobile) ions.
(c) Angular distribution of the axion emission \(g(n_a)=d^2N_a/d\theta_x d\theta_y\) and its \(\theta_x\) projection, with \(\theta_{x,y}=\arctan(p_{x,y}/p_z)\).
(d) Energy spectra of beam electrons and axions for the two ion models.
}}

	\label{fig:source2}
\end{figure}

We study a transverse slice of a uniform electron beam propagating along $+z$ through an initially homogeneous plasma background in a two-dimensional space. 
The plasma is modeled as fully ionized with binary Coulomb collisions neglected.
The periodic \(4\,\mu\mathrm{m}\times4\,\mu\mathrm{m}\) domain contains \(800\times800\) cells, corresponding to \(\Delta x=\Delta y=5\,\mathrm{nm}\), with 15 macroparticles per cell for each species.
The beam has initial energy \(\varepsilon_{be,0}=10~{\rm GeV}\), density \(n_{be}=5\times10^{20}~{\rm cm}^{-3}\), and Lorentz factor \(\gamma_{b0}\simeq1.96\times10^{4}\).
Such high-energy, high-charge electron beams are within the design of FACET-II, where 10-GeV operation is available and a high-charge configuration with bunch charge up to \(5~{\rm nC}\) has been considered \cite{Yakimenko2019FACETII,Emma2025ExtremeBeam,Yakimenko2019NonperturbativeQED}.
The background-electron density is \(n_{pe}=3\times10^{22}~{\rm cm}^{-3}\).
Initial current neutrality is imposed through the return-current drift \( v_{pe,z} = -(n_{be}/n_{pe}) v_{be,z,0}\) where \( v_{be,z,0} = \beta_{b0} c \). Mobile protons with \( Z_i = 1 \),
\( m_i = 1836 m_e \), and \( n_{pi} = n_{pe} + n_{be} \) ensure initial charge neutrality. 
%Unless stated otherwise, stochastic QED photon emission and the associated recoil are utilized. To isolate the role of ion response, we also perform an identical immobile-ion calculation in which the ions are artificially fixed. 
\re{Stochastic QED photon emission is included in all simulations.
To isolate the role of ion response, we also perform an
otherwise identical simulation with immobile ions.} This control preserves the electron-scale current-filamentation dynamics while removing the ion-mediated secondary evolution, which corresponds to the limit \(m_i\to\infty\) in Eq.~\eqref{eq:gamma2} below.
%{
%The simulation results show that current filamentation separates the forward and return currents into dense beam channels anred electron-depleted cavities [Fig.~\ref{fig:source}(a)].
%Their uncompensated longitudinal currents generate azimuthal fields reaching \(B_{\perp,\max}\simeq2\times10^4\,\mathrm{T}\) [Fig.~\ref{fig:source}(b)].}

 %{

Figure~\ref{fig:source} compares the post-saturation filament structures. Both ion models produce electron-scale current filaments with local magnetic fields reaching $B_{\perp,\max}\simeq 2\times10^4\,\mathrm{T}$, corresponding to $\chi_e\sim0.1$ for a 10-GeV electron. Despite comparable peak fields, ion motion markedly alters the filament evolution: neighboring filament channels expand and coalesce with mobile ions [Fig.~\ref{fig:source}(a)], whereas narrow, separated filaments persist with immobile ions [Fig.~\ref{fig:source}(c)]. This difference corresponds to distinct axion-production maps [Figs.~\ref{fig:source}(b) and (d)], showing enhanced emission with mobile ions.

%The axion yield per incident beam electron of the mobile-ion case is \(\eta_a=N_a/N_{be}\simeq0.23g_{ae}^2\).
%For a \(5\)-nC bunch and  \(g_{ae}=10^{-10}\), this corresponds to  \(7.2\times10^{-11}\) axions per shot, 
%\re{
%For the reference mobile-ion simulation, performed with
%\(g_{ae}^{\rm sim}=0.1\), the axion yield per incident beam
%electron at \(t=4.6~\mathrm{ps}\) is
%\(\hat{\eta}_a(g_{ae}^{\rm sim})\equiv N_a/N_e\simeq2.30\times10^{-3}\).
%For an equivalent \(5\)-nC electron bunch, this corresponds
%to approximately \(7.2\times10^7\) axions per shot at the
%simulation coupling,}
\re{
For the reference mobile-ion simulation at
$g_{ae}^{\rm sim}=0.1$, the axion yield per incident
beam electron at $t=4.6~\mathrm{ps}$ is
$\hat{\eta}_a\equiv N_a(g_{ae}^{\rm sim})/N_{be}
\simeq2.30\times10^{-3}$.
For an equivalent $5$-nC electron bunch, this
corresponds to approximately $7.2\times10^7$
axions per shot at the simulation coupling,
about $3.2$ times the immobile-ion yield
[Fig.~\ref{fig:source2}(a)].
In the weak-coupling regime, the yield at an
arbitrary coupling is
$\eta_a(g_{ae})\equiv N_a(g_{ae})/N_{be}
\simeq0.23g_{ae}^2$.} %which is about \(3.2\) times of the immobile-ion yield [Fig.~\ref{fig:source2}(a)]. 
%The axion source is localized near the filament flanks rather than simply at the maxima of beam density or magnetic field [Fig.~\ref{fig:source}].
%
To understand this enhancement, we examine the
temporal evolution of the magnetic-field energy
$\varepsilon_B$ [Fig.~\ref{fig:source2}(a)].
Before $t_s\simeq0.8\,\mathrm{ps}$, both models exhibit nearly identical filamentation growth,
with the magnetic-energy fraction approaching the linear saturation estimate
$\varepsilon_B/\varepsilon_{be,0}\sim\alpha/\gamma_{b0}$, where $\alpha=n_{be}/n_{pe}$.
After saturation, the immobile-ion case evolves slowly, whereas the mobile-ion case undergoes secondary growth governed by
\begin{equation}
\frac{{\varepsilon_B}}{\varepsilon_{be,0}}
\propto e^{2\Gamma_2(t-t_s)},
\qquad
\Gamma_2=\alpha\beta_{b0}
\sqrt{\frac{Z_im_e}{3m_i}}\,\omega_{pe},
\label{eq:gamma2}
\end{equation}
where  $\omega_{pe}$ is the initial plasma frequency~\cite{Peterson2021}.
Consistently, after saturation the mobile-ion density-fluctuation spectrum in [Fig.~\ref{fig:source2}(b)] shifts toward smaller $K_\perp$, while the corresponding mean transverse scale $\lambda_P$ grows substantially faster than in the immobile-ion case. This spectral evolution verifies the expansion and coalescence of the current filaments toward larger transverse structures.
Consequently, comparable values of $B_{\perp,\max}$ can coexist with markedly different magnetic-energy fractions because the mobile-ion field occupies a larger transverse scale.
Notably, the axion-emission rate rises in tandem with this secondary growth [Fig.~\ref{fig:source2}(a)], indicating that ion-mediated filament evolution plays a crucial role in enhancing axion emission.
%Simulations with photon recoil switched off show nearly unchanged magnetic-energy evolution, indicating that radiation reaction does not drive the ion-mediated secondary growth~\cite{SupplementalMaterial1}.
Besides the enhanced yield, the emitted axions exhibit favorable characteristics for downstream detection. The emission is strongly forward collimated, with an angular FWHM of approximately $10$ mrad [Fig.~\ref{fig:source2}(c)]. The axion energy spectrum is concentrated below $1\,\mathrm{GeV}$, with a tail extending to about $2\,\mathrm{GeV}$ [Fig.~\ref{fig:source2}(d)]. %Mobile ions also lead to more pronounced electron cooling, to which enhanced photon emission and the associated recoil contribute.

%, accompanied by stronger electron cooling [Fig.~\ref{fig:source2}(d)].

%To test whether radiation reaction influences the nonlinear magnetic growth, we switch off the QED photon recoil. For mobile ions, the magnetic-energy history is nearly unchanged over the interval shown. For immobile ions, RR gives a moderate late-time enhancement and increases the integrated axion yield by about a factor of \(1.7\). This reflect radiative cooling of long-lived narrow filaments, but it does not account for the ion-mediated secondary growth. We therefore use the mobile-immobile comparison with the same QED radiation model as the primary control for examining the effect of the ion response.

\begin{figure}[t]
\centering
% Artwork update: replace \mathcal H, \varphi_x, f(N_{be}), P_{\rm axion},
% and \mathbf S\cdot\mathbf n by \mathcal H_\perp, \varphi, f_e, W_a,
% and \boldsymbol\zeta\cdot\mathbf b, respectively.
%\includegraphics[width=\columnwidth]{axionfig2.eps}
\begin{overpic}[width=\columnwidth]{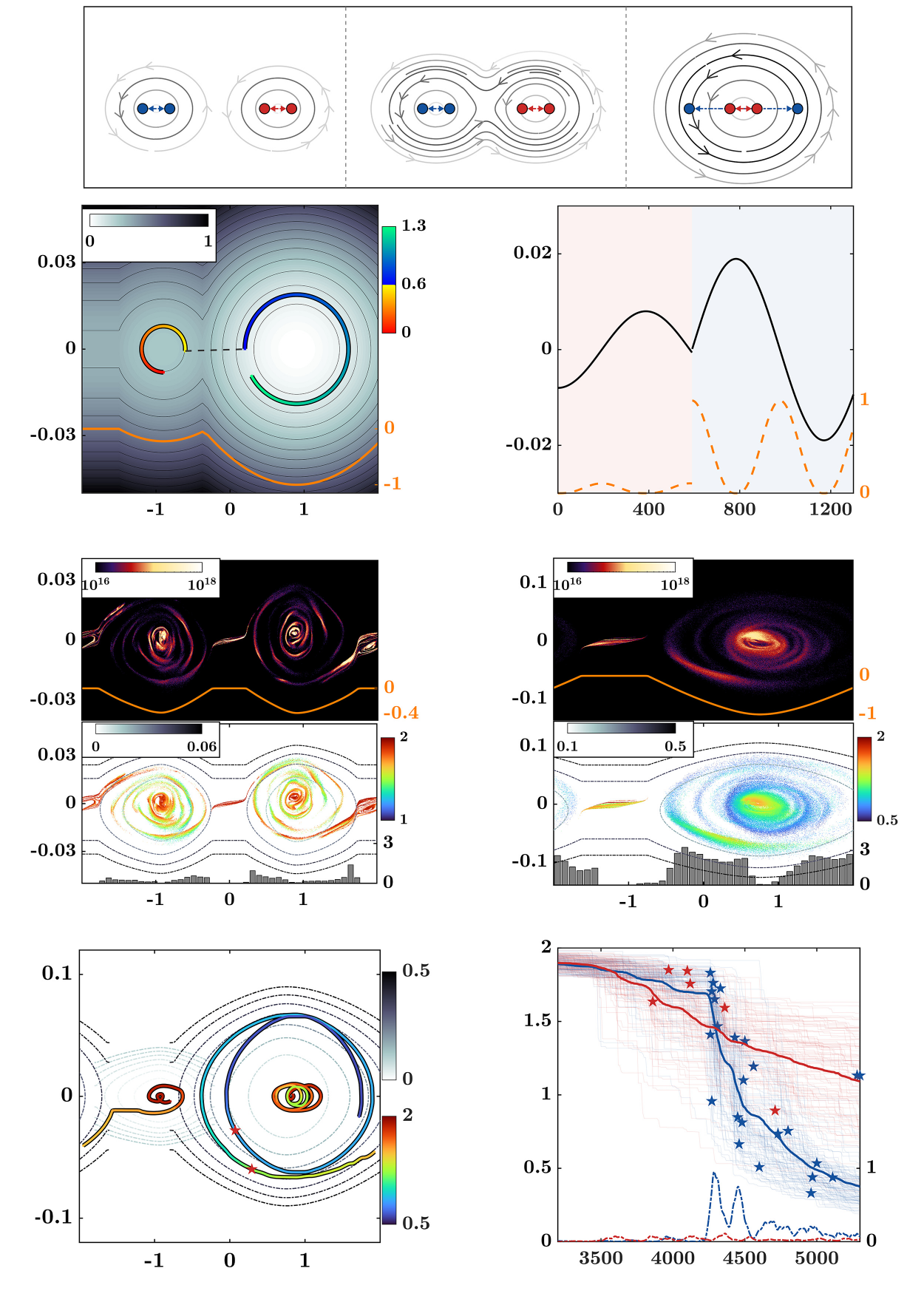}
%\begin{overpic}[width=\columnwidth, grid, tics=10]{axionfig2new.eps}

\newcommand{\labelbox}[6]{%
  \put(#1,#2){%
    \makebox(0,0){%
      \tikz[baseline=(n.base)]{
        \node[
          fill=white,
          fill opacity=#6,
          text opacity=1,
          minimum width=#4,
          minimum height=#5,
          inner xsep=2pt,
          inner ysep=1pt,
          outer sep=0pt,
          rounded corners=1pt
        ] (n) {#3};
      }%
    }%
  }%
}
    
\put(1,97){\textsf{(a)}}
\put(1,83){\textsf{(b)}}
\put(37,83){\textsf{(c)}}
\put(1,58){\textsf{(d)}}
\put(37,58){\textsf{(e)}}
\put(1,28){\textsf{(f)}}
\put(37,28){\textsf{(g)}}

\definecolor{mylabelcolorred}{rgb}{0.80, 0.16, 0.16} 
\definecolor{mylabelcolororige}{rgb}{1, 0.498, 0} 
\definecolor{mylabelcolorblue}{rgb}{0.07, 0.32, 0.62} 
%figa
\put(16.5,97.2){\makebox(0,0){{\scriptsize$\lpmb{\begin{array}{c}\mathrm{nonlinear}\\\mathrm{expansion}\end{array}}$}}}
\put(37.5,98.2){\makebox(0,0){\scriptsize$\lpmb{\mathrm{pre}\text{-}\mathrm{coalescence}}$}}
\put(56.5,98.2){\makebox(0,0){\scriptsize$\lpmb{\mathrm{post}\text{-}\mathrm{coalescence}}$}}
\put(21.5,90){\makebox(0,0){\scriptsize{\textcolor{mylabelcolorred}{$\lpmb{P_x}$}}}}
\put(12,90){\makebox(0,0){\scriptsize{\textcolor{mylabelcolorblue}{$\lpmb{P_x}$}}}}
\put(24,87.5){\makebox(0,0){ $\lpmb{B_{\perp}}$}}
% Coordinate system
\put(6.2,85.5){%
	\begin{tikzpicture}[
		x=1mm,
		y=1mm,
		>=stealth,
		line width=0.6pt
		]

		\draw (0,0) circle (1.05);
		\fill (0,0) circle (0.4mm);   
		
		\draw[->] (1.1,0) -- (4.8,0)
		node[right=-0.2mm] {\scriptsize $\lpmb{x}$};
		
		\draw[->] (0,1.1) -- (0,4.8)
		node[above=-0.2mm] {\scriptsize $\lpmb{y}$};
		
	\end{tikzpicture}%
}

%fig b
\put(18,59){\makebox(0,0){\scriptsize$\lpmb{\xi_x\,[\mu m]}$}}
\put(1.5,74){\makebox(0,0){\rotatebox{90}{\scriptsize$\lpmb{\dot{\xi}_x\,[c]}$}}}
\put(32,67){\makebox(0,0){\scriptsize\rotatebox{90}{\textcolor{mylabelcolororige}{$\boldsymbol{\varphi}_x\,[\mathrm{A.U.}]$}}}}
\put(11.5,81.5){\makebox(0,0){\tiny $\mathcal{H}\,[\mathrm{A.U.}]$}}
\labelbox{27.5}{78.5}{\scriptsize\rotatebox{90}{$\lpmb{t}\,[\lpmb{\mathrm{ps}}]$}}{2mm}{2mm}{0.6}
\put(22.5,67){\makebox(0,0){\scriptsize\textcolor{mylabelcolorblue}{\textnormal{re-injected}}}}
\put(11.5,69.8){\makebox(0,0){\scriptsize\textcolor{mylabelcolorred}{\textnormal{cold-core}}}}

%fig c
\put(54,59){\makebox(0,0){\scriptsize$\lpmb{t}\,[\lpmb{\mathrm{fs}}]$}}
\put(69,66){\makebox(0,0){\tiny\rotatebox{90}{\textcolor{mylabelcolororige}{$\lpmb{W_a}$}}}}
\put(38,74){\makebox(0,0){\rotatebox{90}{\scriptsize$\lpmb{\dot{\xi}_x\,[c]}$}}}
\put(59.5,81.8){\makebox(0,0){\scriptsize\textcolor{mylabelcolorblue}{re-injected}}}
\put(47.5,81.8){\makebox(0,0){\scriptsize\textcolor{mylabelcolorred}{cold-core}}}
%\labelbox{50.5}{65}{%
  %\tiny\textcolor{mylabelcolororige}{$\lpmb{W_{\mathrm{a},2}}=\lpmb{10}\,\lpmb{W_{\mathrm{a},1}}$}%
%}{1mm}{1mm}{0.85}
%\labelbox{55.5}{72}{%
%  \tiny\textcolor{mylabelcolororige}{$\lpmb{10}\,\lpmb{W_{\mathrm{a}}}$}%
%}{1mm}{1mm}{0.85}
%fig d
\put(18,29){\makebox(0,0){\scriptsize$\lpmb{\xi_x\,[\mu m]}$}}
\put(1.5,51){\makebox(0,0){\rotatebox{90}{\scriptsize$\lpmb{\dot{\xi}_x\,[c]}$}}}
\put(1.5,38){\makebox(0,0){\rotatebox{90}{\scriptsize$\lpmb{\dot{\xi}_x\,[c]}$}}}
\put(32.5,48){\makebox(0,0){\scriptsize\rotatebox{90}{\textcolor{mylabelcolororige}{$\boldsymbol{\varphi}_x\,$}}}}
%\put(32.5,40.5){\makebox(0,0){\scriptsize\rotatebox{90}{$\boldsymbol{\gamma}_{\mathrm{e}}\,$}}}

\put(32.5,40){\makebox(0,0){\scriptsize\rotatebox{90}{$\boldsymbol{\gamma}_{\mathrm{e}}\,[10^{4}]$}}}

\put(32.5,34){\makebox(0,0){\scriptsize\rotatebox{90}{$\lpmb{N_a}$}}}
\put(11,42.7){\makebox(0,0){\tiny$\lpmb{\mathcal{H}}$}}
\put(11.2,55.1){\makebox(0,0){\tiny$\lpmb{f_e}$}}

\labelbox{24.2}{55.8}{\scriptsize{$\lpmb{t}=\lpmb{3.4}\,\lpmb{\mathrm{ps}}$}}{1mm}{1mm}{0.75}

%fig e
\put(54,29){\makebox(0,0){\scriptsize$\lpmb{\xi_x\,[\mu m]}$}}
\put(38,39){\makebox(0,0){\rotatebox{90}{\scriptsize$\lpmb{\dot{\xi}_x\,[c]}$}}}
\put(38,51){\makebox(0,0){\rotatebox{90}{\scriptsize$\lpmb{\dot{\xi}_x\,[c]}$}}}
\put(69.5,48){\makebox(0,0){\scriptsize\rotatebox{90}{\textcolor{mylabelcolororige}{$\boldsymbol{\varphi}_x\,$}}}}
\put(69.5,40){\makebox(0,0){\scriptsize\rotatebox{90}{$\boldsymbol{\gamma}_{\mathrm{e}}\,[10^{4}]$}}}
\put(70,34){\makebox(0,0){\scriptsize\rotatebox{90}{$\lpmb{N_a}$}}}
\put(47.5,42.7){\makebox(0,0){\tiny$\lpmb{\mathcal{H}}$}}
\put(47.5,55.1){\makebox(0,0){\tiny$\lpmb{f_e}$}}

\labelbox{60.2}{55.8}{\scriptsize{$\lpmb{t}=\lpmb{4.4}\,\lpmb{\mathrm{ps}}$}}{1mm}{1mm}{0.75}

%f
\put(18,1){\makebox(0,0){\scriptsize$\lpmb{\xi_x\,[\mu m]}$}}
\put(1.5,16){\makebox(0,0){\rotatebox{90}{\scriptsize$\lpmb{\dot{\xi}_x\,[c]}$}}}
\put(33,10.5){\makebox(0,0){\tiny\rotatebox{90}{$\lpmb{\gamma_e\,[10^4]}$}}}
\put(33.2,20.5){\makebox(0,0){\tiny\rotatebox{90}{$\lpmb{\mathcal{H}}\,[\lpmb{\mathrm{A.U.}}]$}}}
\put(23.5,18.8){\makebox(0,0){\scriptsize\textcolor{mylabelcolorred}{$\lpmb{\text{B}}$}}}
\put(11.5,18.8){\makebox(0,0){\scriptsize\textcolor{mylabelcolorblue}{$\lpmb{\text{A}}$}}}

\put(11,12.8){\makebox(0,0){\scriptsize\textcolor{mylabelcolorblue}{$\lpmb{\text{4.2ps}}$}}}
\put(20,7.8){\makebox(0,0){\scriptsize\textcolor{mylabelcolorblue}{$\lpmb{\text{4.3ps}}$}}}
\put(22,12.8){\makebox(0,0){\scriptsize\textcolor{mylabelcolorblue}{$\lpmb{\text{4.8ps}}$}}}
%fig g
\put(54,1){\makebox(0,0){\scriptsize$\lpmb{t}\,[\lpmb{\mathrm{fs}}]$}}
\put(38.5,16){\makebox(0,0){\scriptsize\rotatebox{90}{$\lpmb{\gamma_e\,[10^4]}$}}}
\put(69.5,8){\makebox(0,0){\tiny\rotatebox{90}{{$\lpmb{W_a}$}}}}

{\color{mylabelcolorblue}
  \linethickness{0.8pt}
  \put(44,17){\line(1,0){5}}
}
{\color{mylabelcolorred}
  \linethickness{0.8pt}
  \put(43,14){\line(1,0){5}}
}

\put(49,14){\makebox(0,0){\scriptsize\textcolor{mylabelcolorred}{$\lpmb{\text{B}}$}}}
\put(49,17){\makebox(0,0){\scriptsize\textcolor{mylabelcolorblue}{$\lpmb{\text{A}}$}}}

\end{overpic}
%\caption{Reinjection-enhanced emission. (a) Schematic of nonlinear cavity expansion, and the pre- and post-coalescence stages. (b) Hamiltonian \(\mathcal H_\perp\) in  \((\dot{\xi}_x,\xi_x)\), with the cold-core orbit reinjected trajectories (time-colored), and effective potential \(\varphi\) (orange).
%(c) Time evolution of \(\dot{\xi}_x\) (black) and the spin-averaged axion-emission rate \(W_a=dP_a/dt\) for the two orbit classes.
%(d) and (e) Beam-electron phase-space density \(f_e(\dot{\xi}_x,\xi_x)\) (upper) and mean Lorentz factor \(\langle\gamma_e\rangle\,(10^4)\) (lower) at \(t=3.4\) and \(4.4\,\mathrm{ps}\), respectively. Orange curves, gray dashed contours, and gray bars show \(\varphi\), \(\mathcal H_\perp\), and the accumulated axion source projected onto \(\xi_x\), respectively.
%(f) Representative reinjected (\(A\)) and cold-core (\(B\)) trajectories, colored by \(\gamma_e\). (g) Ensemble-averaged Lorentz factor \(\langle\gamma_e\rangle\)
%(solid) and axion-emission rate (dashed) for \(A\) (blue) and \(B\) (red). Stars in (f)(g) mark axion emission events.}

\caption{%Reinjection-enhanced emission.
(a) Schematic of nonlinear filament expansion and the pre- and post-coalescence stages.
(b) Hamiltonian \(\mathcal H_\perp\) in \((\dot{\xi}_x,\xi_x)\), with the
cold-core and reinjected trajectories (time-colored), and effective potential
\(\varphi\) (orange).
(c) Time evolution of \(\dot{\xi}_x\) (black) and the spin-averaged
axion-emission rate \(W_a=dP_a/dt\) for the two orbit classes.
(d) and (e) Beam-electron phase-space density
\(f_e(\dot{\xi}_x,\xi_x)\) (upper) and mean Lorentz factor
\(\langle\gamma_e\rangle\) (lower) at
\(t=3.4\) and \(4.4\,\mathrm{ps}\), respectively.
Orange curves and gray dashed contours show \(\varphi\) and
\(\mathcal H_\perp\), respectively. Gray bars show the axion source
generated during the preceding \(100~\mathrm{fs}\) time window,
i.e., within \([t-100~\mathrm{fs},t]\), and projected onto \(\xi_x\).
(f) Representative reinjected (\(A\)) and cold-core (\(B\)) trajectories,
colored by \(\gamma_e\).
(g) Ensemble-averaged Lorentz factor \(\langle\gamma_e\rangle\) (solid)
and axion-emission rate (dashed) for \(A\) (blue) and \(B\) (red).
Stars in (f) and (g) mark axion-emission events.}

\label{fig:mechanism}
\end{figure}

To identify the dynamics of radiating electrons, we consider \(x_f(t)\) as the filament center and define the co-moving coordinate \(\xi_x=x-x_f(t)\) with \(\dot\xi_x=v_x-\dot x_f\). Assuming \(v_z\simeq c\), \(\gamma_e\simeq\bar\gamma_{\rm orb}\) over one transverse orbit, and slowly moving filament, the electron transverse dynamics is characterized by
\begin{equation}
	\mathcal H_\perp(\dot{\xi}_x, \xi_x)=
	\frac{\dot{\xi}_x^2}{2}+\varphi(\xi_x),
	\qquad
	\frac{d\varphi}{d\xi_x}=
	-\frac{ecB_y}{\bar{\gamma}_{\rm orb}m_e},
	\label{eq:hamiltonian}
\end{equation}
where \(\varphi\) is the effective potential. For a quasistatic filament, the Hamiltonian \(\mathcal H_\perp\) is nearly conserved. Near the center, \(B_y\simeq-B'_0\xi_x\) gives \(\varphi\simeq\Omega^2\xi_x^2/2\), with \(\Omega^2=ecB'_0/(\bar\gamma_{\rm orb}m_e)\). The inner contours correspond to weakly radiating cold-core orbits [Fig.~\ref{fig:mechanism}(b)]. 
During coalescence, \(\varphi(\xi_x,t)\) becomes time dependent and \(d\mathcal H_\perp/dt\simeq\partial\varphi/\partial t\neq0\). 
As the barrier between neighboring wells falls, some interior core electrons cross the instantaneous separatrix and then enter other extended orbits [Figs.~\ref{fig:mechanism}(d)--\ref{fig:mechanism}(f)]. 
These electrons repeatedly traverse the strong-field flanks, where \(\langle\chi_e\rangle\) rises from about \(0.035\) before coalescence to \(0.1\) afterward, resulting in substantially enhanced axion-emission rates [Figs.~\ref{fig:mechanism}(c) and \ref{fig:mechanism}(g)]. %Spin changes the rate quantitatively but not the ordering of the two orbit classes.
%
%For reinjected orbits spanning a merged filament, \gr{\(r_\beta\sim r_f\)}, the characteristic emission angle approximates
%\begin{equation}
%\theta_{a,\mathrm{char}}
%\sim\frac{\Omega r_\beta}{c}
%\sim\left(
%\frac{eB_{\rm fil}r_f}{\bar{\gamma}_{\rm orb}m_ec}
%\right)^{1/2}.
%\label{eq:angle_scaling}
%\end{equation}
%where \(B_{\rm fil}=B'_0r_f\). 
Let \(r_\beta\) denote the transverse orbital amplitude relative to the
filament center and \(r_f\) the characteristic transverse radius of the
merged filament. For reinjected electrons whose trajectories span the
filament, \(r_\beta\sim r_f\). The corresponding characteristic emission
angle can then be estimated as
\begin{equation}
\theta_{a,\mathrm{char}}
\sim\frac{\Omega r_\beta}{c}
\sim\left(
\frac{eB_{\rm fil}r_f}{\bar{\gamma}_{\rm orb}m_ec}
\right)^{1/2}.
\label{eq:angle_scaling}
\end{equation}
Here \(B_{\rm fil}=B'_0r_f\) denotes the characteristic magnetic field
at the filament flank.
With \(B_{\rm fil}\simeq2\times10^4\,\mathrm{T}\), \(r_f\simeq0.1\text{--}0.3\,\mu\mathrm{m}\), and \(\bar\gamma_{\rm orb}\simeq1.96\times10^4\), Eq.~\eqref{eq:angle_scaling} gives \(8\text{--}13\,\mathrm{mrad}\), consistent with Fig.~\ref{fig:source2}(c).

A high-resolution 1D PIC simulation elucidates the phase-space process. 
%The \(4\,\mu\mathrm{m}\) domain uses \(\Delta x=1\,\mathrm{nm}\) and 100 macroparticles per cell per species. 
At \(3.4\,\mathrm{ps}\), two potential wells contain compact cold cores [Fig.~\ref{fig:mechanism}(d)]. 
By \(4.4\,\mathrm{ps}\), the wells have merged and part of the core population occupies extended orbits [Fig.~\ref{fig:mechanism}(e)]. The projected axion source peaks on the potential flanks.
A tracked electron from ensemble \(A\) is detrapped during coalescence and crosses these flanks, whereas ensemble \(B\) remains near the potential minimum [Fig.~\ref{fig:mechanism}(f)]. The sampled axion events and ensemble rates in Fig.~\ref{fig:mechanism}(g) follow the similar tendency. 
\bl{Thus the ion-mediated nonlinear stage enhances emission through the combined increase of the magnetic scale size and the transfer of electrons onto the position with strong magnetic fields.}

\begin{figure}[t]
\centering
% Artwork update: use B_{\perp,\rm char}; label g_{ae} as dimensionless;
% and correct the panel-(b) n_{be} axis multiplier.
%\includegraphics[width=\columnwidth]{axionfig32.eps}
\begin{overpic}[width=\columnwidth]{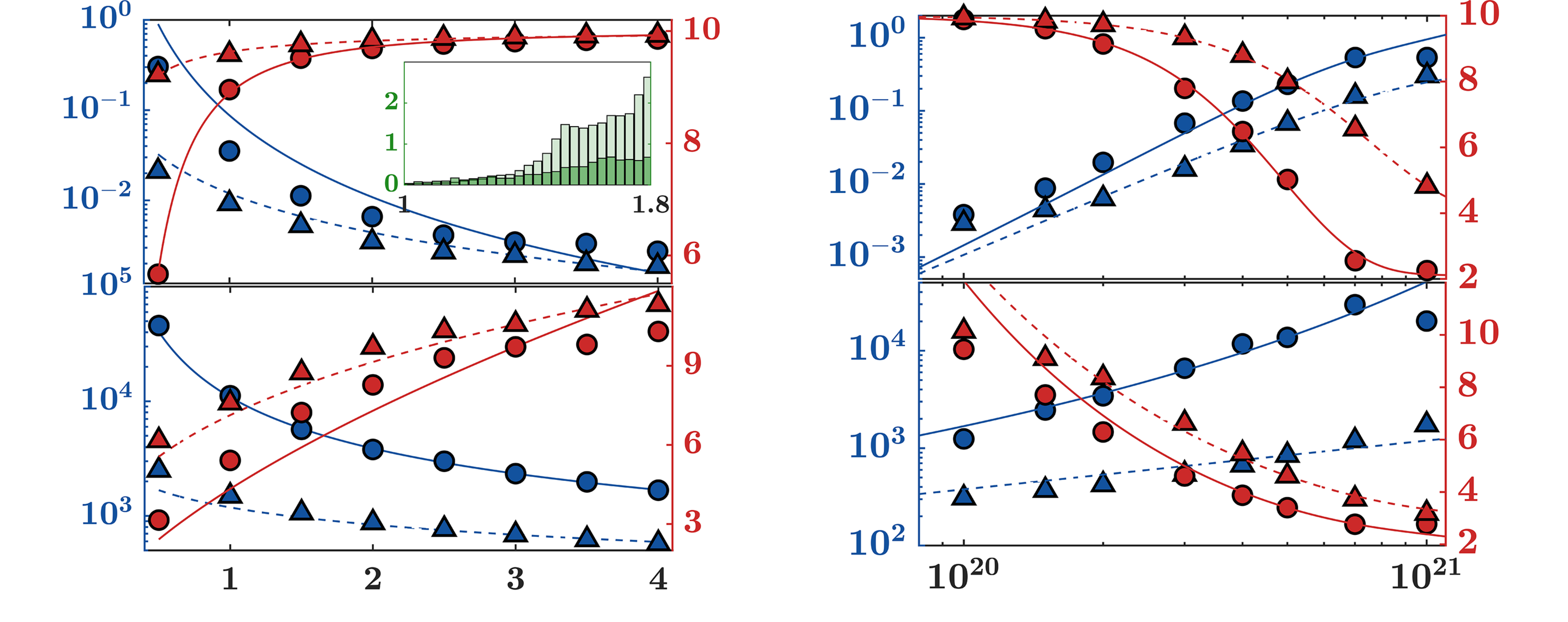}
%\begin{overpic}[width=\columnwidth, grid, tics=10]{axionfig32.eps}	
\put(-1,38){\textsf{(a)}}
\put(49,38){\textsf{(b)}}
\definecolor{mylabelcolorblue}{rgb}{0.07, 0.32, 0.62} 
\definecolor{mylabelcolorrad}{rgb}{0.80, 0.16, 0.16} 
\definecolor{mylabelcolorgreen}{rgb}{0.13, 0.55, 0.13} 
%a	
\put(26,0){\makebox(0,0){\scriptsize{$\lpmb{n_{pe}\,[10^{22}\,\mathrm{cm}^{-3}]}$}}}
\put(1.5,31){\makebox(0,0){\scriptsize{\rotatebox{90}{\textcolor{mylabelcolorblue}{$\lpmb{\hat{\eta}_{a}}$}}}}}
\put(2.5,14){\makebox(0,0){\scriptsize{\rotatebox{90}{\textcolor{mylabelcolorblue}{$\lpmb{B_{\perp}\,[\lpmb{\mathrm{T}}]}$}}}}}
\put(46.5,30.5){\makebox(0,0){\scriptsize{\rotatebox{90}{\textcolor{mylabelcolorrad}{$\lpmb{\bar{\varepsilon}_{be}\,[\mathrm{GeV}]}$}}}}}
\put(46.5,13.5){\makebox(0,0){\scriptsize{\rotatebox{90}{\textcolor{mylabelcolorrad}{$\lpmb{g_{ae}\,[10^{-6}]}$}}}}}

\put(33,26){\makebox(0,0){\tiny{$\lpmb{t\,[ps]}$}}}
\put(21,33){\makebox(0,0){\tiny{\textcolor{mylabelcolorgreen}{$\lpmb{\frac{dN_a}{dt}}$}}}}
\put(30,31){\makebox(0,0){\tiny{\textcolor{mylabelcolorgreen}{$\lpmb{\times 5}$}}}}
%b
\put(77,0){\makebox(0,0){\scriptsize{$\lpmb{n_{be}\,[\,\mathrm{cm}^{-3}]}$}}}
\put(51,31){\makebox(0,0){\scriptsize{\rotatebox{90}{\textcolor{mylabelcolorblue}{$\lpmb{\hat{\eta}_a}$}}}}}
\put(52,14){\makebox(0,0){\scriptsize{\rotatebox{90}{\textcolor{mylabelcolorblue}{$\lpmb{B_{\perp}\,[\lpmb{\mathrm{T}}]}$}}}}}
\put(98,31.5){\makebox(0,0){\scriptsize{\rotatebox{90}{\textcolor{mylabelcolorrad}{$\lpmb{\bar{\varepsilon}_{be}\,[\mathrm{GeV}]}$}}}}}
\put(98.31,13.5){\makebox(0,0){\scriptsize{\rotatebox{90}{\textcolor{mylabelcolorrad}{$\lpmb{g_{ae}\,[10^{-6}]}$}}}}}
\end{overpic}
\caption{Parameter scans calculated with \(g_{ae}^{\rm sim}=0.1\).
	(a) Background-density $n_{pe}$ scan at fixed
	\(n_{be}=5\times10^{20}\,\mathrm{cm}^{-3}\) and \(L_p=560\,\mu\mathrm{m}\)
	(\re{the interaction length}):
	reference-coupling axion yield $\hat{\eta}_a$, final mean electron energy,
	transverse magnetic field, and \(N_{\rm pair}=1\) coupling reach.
	The inset compares mobile and immobile ions at
	\(n_{pe}=5\times10^{21}\,\mathrm{cm}^{-3}\).
	(b) Beam-density $n_{be}$ scan at fixed \(\alpha=1/60\) and
	\(L_p=1400\,\mu\mathrm{m}\).
	Here, circles and triangles denote mobile- and immobile-ion cases, respectively, with the solid and dashed corresponding to the predicted scalings.
}
\label{fig:scan}
\end{figure}

Figure~\ref{fig:scan} compares the simulated
density dependence with the leading scaling estimates. Using the LCFA scaling
$W_a/g_{ae}^2\propto\chi_e^p/\gamma_e$,
with $p\simeq2.69$, the characteristic magnetic
field and reference-coupling axion yield scale as
\begingroup
\begin{equation}
	\begin{aligned}
		B_{\perp,\mathrm{char}}
		&\propto
		\begin{cases}
			n_{pe}^{-1/2}e^{\Gamma_2\Delta t_{\rm nl}},
			& n_{be}\ \mathrm{fixed},\\
			n_{be}^{1/2}e^{\Gamma_2\Delta t_{\rm nl}},
			& \alpha\ \mathrm{fixed},
		\end{cases}
		\\[3pt]
		\hat{\eta}_a
		&\propto
		B_{\perp,\mathrm{char}}^{\,p-2}
		\left(
		\gamma_{b0}^{p-2}-\gamma_f^{p-2}
		\right).
	\end{aligned}
	\label{eq:scan_scaling}
\end{equation}
\endgroup
	%Here \(\Gamma_2\) is given by Eq.~\eqref{eq:gamma2}, while
	%\(\gamma_f\) is obtained from a simple fit to the final mean beam electron energy.
	%The axion-yield scaling follows by combining the emission rate with
	%the leading radiative-cooling scaling. The fitting forms and intermediate
	%steps are given in the Supplemental Material.
\re{Here \(\Gamma_2\) is given by Eq.~\eqref{eq:gamma2}, \(\Delta t_{\rm nl}\) is the duration of the nonlinear stage, and \(\gamma_{b0}\) and \(\gamma_f\) denote the initial and final mean beam-electron Lorentz factors, respectively; the latter is obtained from a simple fit to the final mean beam-electron energy. The axion-yield scaling follows by combining the emission rate with the leading radiative-cooling scaling. The fitting forms and intermediate steps are given in the Supplemental Material.}    
%\re{Here $B_{\perp,\mathrm{char}}$ denotes the characteristic
%transverse magnetic-field amplitude estimated from
%the filament current and electron skin-depth scaling,
%including the ion-mediated secondary growth. $\Delta t_{\rm nl}$ is the duration of the nonlinear stage,
%while $\gamma_{b0}$ and $\gamma_f$ denote the
%initial and final mean beam-electron Lorentz factors,
%respectively. The latter is obtained from a fit to
%the simulated final mean electron energy.
%The intermediate steps and fitting procedure are
%given in the Supplemental Material.}
%
For Fig.~\ref{fig:scan}(a), fixed \(n_{be}\) gives \(\Gamma_2\propto n_{pe}^{-1/2}\); increasing \(n_{pe}\) therefore weakens the field, reduces the axion yield, and suppresses beam cooling. In the immobile-ion limit, \(m_i\to\infty\) gives \(\Gamma_2\to0\), leaving the leading \(B_{\perp,\mathrm{char}}\propto n_{pe}^{-1/2}\) dependence. At \(n_{pe}=3\times10^{22}\,\mathrm{cm}^{-3}\), ion motion raises the integrated yield by about \(3.2\), while at \(n_{pe}=5\times10^{21}\,\mathrm{cm}^{-3}\) the enhancement reaches about \(14.5\) [inset of Fig.~\ref{fig:scan}(a)]. For Fig.~\ref{fig:scan}(b), fixed \(\alpha=1/60\) gives \(\Gamma_2\propto n_{be}^{1/2}\), so increasing \(n_{be}\) strengthens the field and axion yield while lowering the final beam energy.

\begin{figure}[t]
\centering
\begin{overpic}[width=0.85\columnwidth]{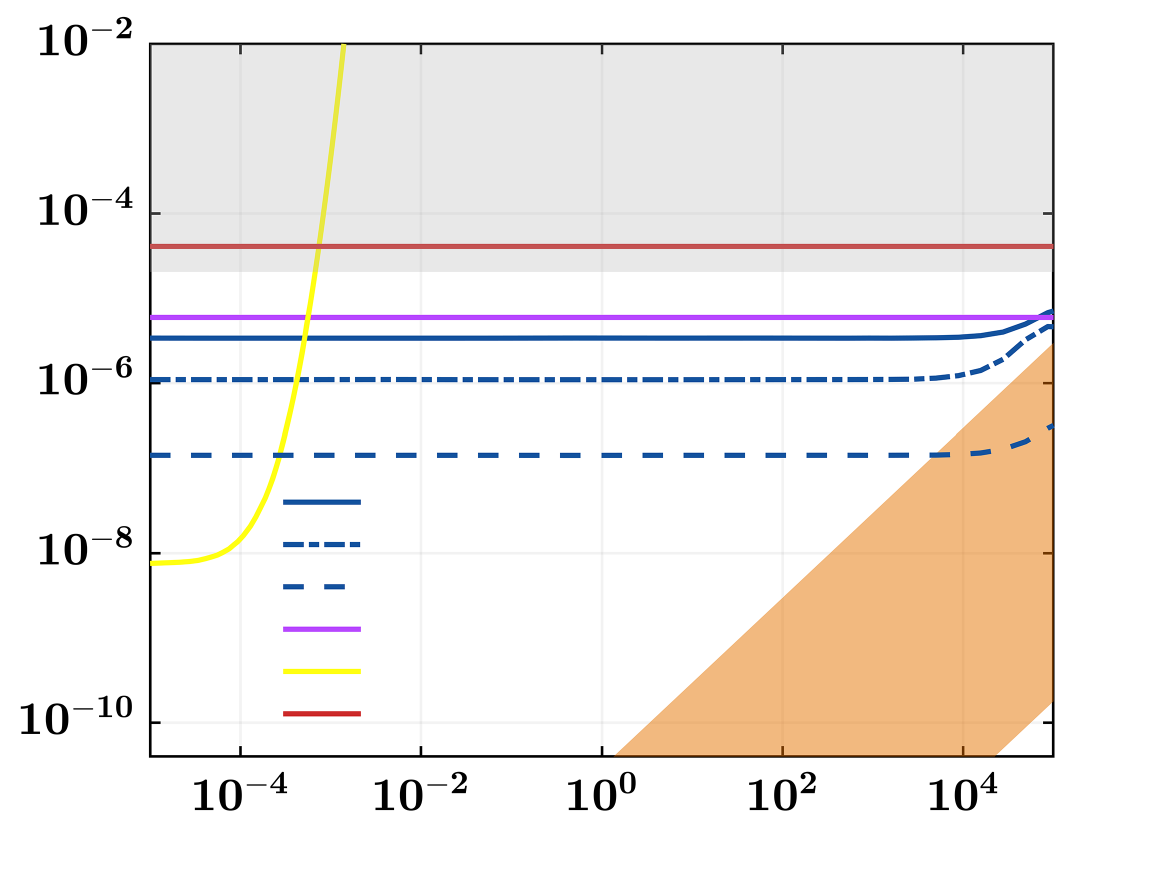}
%\begin{overpic}[width=0.85\columnwidth, grid, tics=10]{axionfig4.eps}
\put(50,1){\makebox(0,0){\normalsize $\lpmb{m_a\,[\lpmb{\mathrm{eV}}]}$}}
\put(-1,38){\makebox(0,0){\rotatebox{90}{{\large $\lpmb{g_{ae}}$}}}}

\put(38.5,31.5){\makebox(0,0){\tiny{\fontseries{b}\selectfont FACET-II}}}
\put(37,27.8){\makebox(0,0){\tiny{\fontseries{b}\selectfont SHINE}}}
\put(39.9,24.2){\makebox(0,0){\tiny{\fontseries{b}\selectfont Future-beam}}}
\put(36.6,20.7){\makebox(0,0){\tiny{\fontseries{b}\selectfont NA-64}}}
\put(37.4,16.9){\makebox(0,0){\tiny{\fontseries{b}\selectfont Terrano}}}
\put(39.1,13.2){\makebox(0,0){\tiny{\fontseries{b}\selectfont LAE-Axion}}}

\put(78.6,20.2){\makebox(0,0){\normalsize$\lpmb{\mathrm{DFSZ}}$}}
\put(68.6,60.2){\makebox(0,0){\normalsize$\lpmb{\mathrm{Electron\ g\text{-}2}}$}}

% 在 (40.5, 31.5) 处画一个 3mm 宽、2mm 高的空心矩形
\put(22.5,11.5){\framebox(26,22){}}
\end{overpic}
\caption{Projected \(N_{\rm pair}=1\) reach from axion-induced pair production in a segmented tungsten calorimeter. The solid, dash-dotted, and dashed blue curves represent the FACET-II, SHINE, and future-beam scenarios, respectively. The red solid curve shows the LAE-Axion~\cite{Vacalis2025LAE} projection, while the magenta and yellow curves denote the NA64 and Terrano 2015~\cite{Terrano2015} constraints. The gray region indicates the electron-\(g-2\)~\cite{Yan2019ExoticSpin} constraint, and the light-orange band represents the DFSZ parameter space.}
\label{fig:reach}
\end{figure}

The forward axions can be detected through external pair production, \(a+{}^{A}_{Z}X\to{}^{A}_{Z}X+e^+e^-\), in a nuclear Coulomb field~\cite{AriasAragon2025Pair}. In the light-axion limit, \(m_a\ll m_e\),
\begin{equation}
\sigma_{aee}(E_a)\simeq
2.6\frac{g_{ae}^2}{4\pi\alpha_{\rm fs}}\,
\sigma_{\gamma ee}(E_a),
\label{eq:pair_cross_section}
\end{equation}
where \(\alpha_{\rm fs}\) is the fine-structure constant and \(\sigma_{\gamma ee}\) is the Bethe--Heitler pair-production cross section. We take a segmented tungsten--scintillator calorimeter with cumulative tungsten thickness \(L_W=10\,\mathrm{m}\) as a benchmark~\cite{McNabb2009,CALICE2014,CALICE2018,AriasAragon2025Pair}.
The expected pair count is
%\begin{equation}
%	N_{\rm pair}
%	=
%	f_{\rm rep}T\,N_{be}^{\rm bunch}\eta_a
%	\left(\frac{g_{ae}}{g_{ae}^{\rm sim}}\right)^2
%	n_WL_W\langle\sigma_{aee}\rangle_a ,
%	\label{eq:events}
%\end{equation}

\begin{equation}
N_{\rm pair}
=
f_{\rm rep}T N_{be}^{\rm bunch}
\eta_a(g_{ae})\,
n_WL_W\langle\sigma_{aee}\rangle_a .
\label{eq:events}
\end{equation}
%{
where \(f_{\rm rep}\) is the bunch repetition rate, \(T\) the running time, \(N_{be}^{\rm bunch}\) the bunch population, $n_W$ is the atomic number density of tungsten, and \(\langle\sigma_{aee}\rangle_a\) the spectrum-averaged cross section. Since production and detection each contribute one factor of \(g_{ae}^2\), the scan curves with fixed bunch population obey \(g_{ae}\propto\hat\eta_a^{-1/4}\).
Figure~\ref{fig:reach} shows the \(N_{\rm pair}=1\) projections for \(L_W=10\,\mathrm{m}\). A FACET-II-like \(10\)-GeV, \(5\)-nC bunch at \(30\,\mathrm{Hz}\), using the reference yield \(\eta_a=0.23g_{ae}^2\), reaches \(g_{ae}^{(1)}\simeq3.4\times10^{-6}\) in seven days~\cite{FACETIICDR,Yakimenko2019FACETII,Emma2025ExtremeBeam}. For SHINE, an \(8\)-GeV, \(100\)-pC bunch at \(1\,\mathrm{MHz}\) with \(\eta_a=7.8\times10^{-3}g_{ae}^2\) reaches \(1.1\times10^{-6}\) in 30 days~\cite{Huang2023SHINE}. A future \(1\)-MHz beam with \(n_{be}=10^{21}\,\mathrm{cm}^{-3}\) extends the projected low-mass reach to the \(10^{-8}\) range. The projected sensitivities suggest that all three scenarios could extend the reach beyond the electron-$g-2$, LAE-Axion, and NA64 benchmarks in the low-mass regime, while the Terrano 2015 constraint remains more stringent at ultralight masses.

%

%Current filamentation can therefore serve as a self-organized source of electron-coupled axions. {\color{red}The immobile-ion control shows that comparable local peak fields can coexist with different magnetic scales and axion yields. Ion response enhances the source through both larger-scale magnetic structures and coalescence-driven phase-space transport, which places part of the electron population on strongly radiating orbits.} Radiation reaction modifies the immobile-ion baseline but is subdominant in the mobile-ion nonlinear growth over the simulated interval. The resulting axion beam has a \(\sim10\)-mrad divergence and, with the benchmark tungsten converter, gives projected \(g_{ae}\) reaches at the \(10^{-6}\) level for FACET-II and SHINE repetition rates. The same mechanism may also be relevant wherever relativistic current filamentation develops in collisionless plasmas~\cite{MedvedevLoeb1999,Spitkovsky2008Shock,Gong2023Slingshot}.
%Our results not only high light that the ion dynamics can determine strong magnetic fields of plasma instabilities to emit weakly coupled particles, but also indicate the ubiquity of electron-coupled particle emission in astrophysics because ion-Weibel filamentation is common in relativistic astrophysical environments
\bl{In summary, we have demonstrated that current filamentation, often treated as a limitation on relativistic-beam transport, can act as a self-organized axion source. 
The decisive ingredient is ion-mediated nonlinear evolution: ion motion expands and merges the current filament channels, strengthening the magnetic field and transferring cold-core electrons onto large-amplitude radiating orbits. 
This collective phase-space transport sets the source brightness and produces a forward axion beam with a divergence of about 10 mrad. When coupled to a tungsten pair calorimeter, the source gives projected single-event sensitivities near \(g_{ae} = 10^{-6}\) for FACET-II and SHINE repetition rates. 
Our results demonstrate that ion-mediated filament evolution can regulate axion emission by restructuring electron trajectories in self-generated fields. Similar mechanisms may also be relevant to relativistic astrophysical plasmas in which Weibel-type filamentation develops~\cite{MedvedevLoeb1999,Spitkovsky2008Shock,Gong2023Slingshot}.}

\begin{acknowledgments}
This work is supported by the National
Natural Science Foundation of China (Grants Nos.12474312
and 12574377), and the National Key R\&D Program of China
(Grant No. 2021YFA1601700). Z.G. acknowledges the support from the Strategic Priority Research Program of the Chinese Academy of Sciences (Grant No. XDB1550100), the National Key R\&D Program of China (2025YFF0515103), the CAS Project for Young Scientists in Basic Research (Grant No. YSBR-141), the National Natural Science Foundation of China (Grant No. 12675316 and 12447101), and the support from the HPC Cluster of ITP-CAS.
The EPOCH code used in this work was in part funded by the UK EPSRC grants EP/G054950/1, EP/G056803/1, EP/G055165/1, EP/ M022463/1 and EP/P02212X/1.
\end{acknowledgments}

\bibliography{axion_references}

\end{document}